\documentclass[a4paper,11pt]{article}
\usepackage{paper}
\usepackage{fontawesome5}
\usepackage{algorithm}
\usepackage{algorithmicx}
\usepackage{algpseudocode}

\pdfoutput=1

\hypersetup{
    pdftitle={Distributed Generalized Linear Models: A Privacy-Preserving Approach},
    pdfauthor={Daniel Tinoco, Raquel Menezes, Carlos Baquero}
}

\newcommand{\bib}{bibliography.bib}

\begin{document}

% Enter title:
\title{Distributed Generalized Linear Models: A Privacy-Preserving Approach}

% Enter authors:
\author{Daniel Tinoco, Raquel Menezes, Carlos Baquero
%
% Enter affiliations and acknowledgements:
\thanks{D. Tinoco (\hspace{1px}\faEnvelope[regular]\hspace{1px})\\
Centro de Matemática, Universidade do Minho, Braga, Portugal\\
DEI-FEUP \& INESC TEC, Universidade do Porto, Porto, Portugal\\
E-mail: danieltinoco@fe.up.pt\\\\
R. Menezes\\
Centro de Matemática, Universidade do Minho, Braga, Portugal\\
E-mail: rmenezes@math.uminho.pt\\\\
C. Baquero\\
DEI-FEUP \& INESC TEC, Universidade do Porto, Porto, Portugal\\
E-mail: cbm@fe.up.pt
}
}

% Enter date:
\date{March 2025}   

% Enter permanent URL (can be commented out):
%\available{https://github.com/}

\begin{titlepage}
\maketitle

% Enter abstract:
This paper presents a novel approach to classical linear regression, enabling model computation from data streams or in a distributed setting while preserving data privacy in federated environments. We extend this framework to generalized linear models (GLMs), ensuring scalability and adaptability to diverse data distributions while maintaining privacy-preserving properties. To assess the effectiveness of our approach, we conduct numerical studies on both simulated and real datasets, comparing our method with conventional maximum likelihood estimation for GLMs using iteratively reweighted least squares. Our results demonstrate the advantages of the proposed method in distributed and federated settings.

\end{titlepage}

% Enter main text:
\section{Introduction}
\label{sec1}

Linear models are among the most fundamental tools in statistical and machine learning \citep{pml1Book}, providing a simple yet powerful framework for modeling relationships between a response variable and one or more predictor variables. In traditional centralized approaches, model fitting requires aggregating all relevant data at a single location, where statistical techniques are applied to estimate the parameters that best explain the relationship between predictor and response variables. However, this approach is often impractical in modern applications where data is distributed across multiple sources, privacy constraints prevent direct data sharing \citep{10.1145/3460427}, and new observations continuously arrive in a streaming fashion \citep{10.2307/2347147}.

A key challenge in such settings is how to compute and share model results efficiently while ensuring that sensitive data remains private. Federated learning \citep{McMahanMRHA17} has emerged as a promising paradigm, allowing models to be trained across decentralized data sources without exposing raw data. This is particularly relevant in domains such as healthcare, finance, and mobility analytics, where privacy concerns restrict data access. Additionally, the computational burden of retraining models from scratch when new data becomes available is often prohibitive, highlighting the need for scalable methods that support incremental updates.

To address these challenges, we propose alternative approaches for fitting linear models that make them more suitable for distributed and privacy-preserving environments. Some of the ideas presented here are inspired by classic works, such as \cite{10.2307/2347147}, which discusses solutions for large or weighted linear least squares problems, and \cite{10.2307/2284221} dedicated to regression updating.
Specifically, we introduce a modified ordinary least squares (OLS) method with QR decomposition, where the cross-product of $\mathbf{Q}$ transposed and the response variable are computed within the augmented matrix $\mathbf{R}$. This method enables efficient model computation from data streams while reducing memory requirements.

While linear models are widely used, they rely on the assumption of a linear relationship between variables and normally distributed errors, which may not always hold in practice. Generalized Linear Models (GLMs) extend linear models by allowing the response variable to follow a broader class of probability distributions and incorporating a non-linear transformation of the predictors. In this work, we demonstrate how the computation of GLMs can be reformulated as an iteratively distributed estimation of multiple OLS solutions, enabling efficient fitting in decentralized settings.

In addition to the theoretical and algorithmic contributions, we conduct numerical studies on both simulated and real datasets to evaluate the performance of our methods. We assess their accuracy, computational efficiency, and robustness in distributed and federated settings, where floating-point arithmetic effects can significantly impact attainable precision \citep{8766229}. The results highlight the potential of our approach for scalable and privacy-preserving model estimation in modern data-driven applications.\\

\section{Efficient and distributed linear regression models}
\subsection{Linear regression}
\label{sec:lm}
Linear regression models describe a numeric response variable $y$ as a linear combination of predictor variables $x_k$, where $k = 1, \dots, p-1$, assuming homoscedasticity \citep{RWC03}. Each of these variables is observed across $n$ instances. The fitted values are obtained by summing the coefficients $\beta_k$ multiplied by each $x_{ik}$, with $i = 1, \dots, n$, along with an intercept $\beta_0$ \citep{Arnold2019-ug}, expressed by the formula:
\begin{equation}
y_i = \beta_0 + \beta_1 x_{i1} + \dots + \beta_{p-1} x_{ip-1} + \varepsilon_i
\label{eq:lm}
\end{equation}

This linear model \eqref{eq:lm} can also be expressed in matrix notation, following Björck's notation \citep{Bjorck1996-gj}, using an $n \times 1$ vector of observations $\mathbf{Y}$ and a $n \times p$ design matrix $\mathbf{X}$:
\begin{equation} \mathbf{Y} = \mathbf{X} \boldsymbol\beta + \boldsymbol\varepsilon \end{equation}

Among the various methods available for fitting linear models, the least squares approach \citep{10.1093/biomet/54.1-2.1} is commonly employed. This method determines the $p \times 1$ vector of regression coefficients $\boldsymbol\beta$ by minimizing the sum of squared residuals, where the residual vector $\mathbf{r}$ is defined as $\mathbf{r} = \mathbf{Y} - \mathbf{X}\boldsymbol{\hat{\beta}}$. This minimization leads to the normal equations:
\begin{equation} 
\mathbf{X}^\mathsf{T}\mathbf{X}\boldsymbol\beta = \mathbf{X}^\mathsf{T}\mathbf{Y}
\label{eq:normal_lm}
\end{equation}

The most common solution to this problem is the ordinary least squares (OLS) estimator, given by:
\begin{equation}
\boldsymbol{\hat{\beta}} = (\mathbf{X}^\mathsf{T}\mathbf{X})^{-1} \mathbf{X}^\mathsf{T} \mathbf{Y} \label{eq:lm_beta_hat}
\end{equation}

However, solving this equation numerically can be challenging due to the often ill-conditioned\footnote{A small change in the constant coefficients can result in a large change in the solution.} nature of the cross-product matrix $\mathbf{X}^\mathsf{T}\mathbf{X}$, which may lead to numerical instability. Furthermore, the computational cost of matrix inversion poses a significant constraint, particularly when handling large data sets or ill-conditioned matrices \citep{10.1093/imamat/12.3.329}.\\

\subsection{QR decomposition}
\label{sec:qr}

An alternative to solving the least squares problem without explicitly computing the ill-conditioned cross-product matrix $\mathbf{X}^\mathsf{T}\mathbf{X}$ is QR decomposition. Given $\mathbf{X} \in \mathbb{R}^{n \times p}$ (or $\mathbf{X} \in \mathbb{C}^{n \times p}$) with $n \geq p$, the decomposition expresses $\mathbf{X}$ as:
\begin{equation}
\mathbf{X} = \mathbf{Q}\mathbf{R}
\label{eq:qr}
\end{equation}
where $\mathbf{Q}$ is an $n \times p$ matrix with orthonormal columns ($\mathbf{Q}^\mathsf{T} = \mathbf{Q}^{-1}$), and $\mathbf{R}$ is a $p \times p$ upper triangular matrix. 

 QR decomposition can be achieved through various methods, such as Gram–Schmidt orthogonalization, with the modified Gram–Schmidt process being the most commonly used version, Householder transformation, or employing Givens rotations, each with its own set of advantages and disadvantages \citep{10.5555/248979}.

\subsubsection{Solving the least squares problem}
Substituting \eqref{eq:qr} into the normal equations yields:
\begin{equation}
\mathbf{R}^\mathsf{T} \mathbf{R} \boldsymbol\beta = \mathbf{R}^\mathsf{T} \mathbf{Q}^\mathsf{T} \mathbf{Y}
\label{eq:qr_change}
\end{equation}
where $\mathbf{R}$ is nonsingular if $\mathbf{X}^\mathsf{T}\mathbf{X}$ is. This allows solving for $\boldsymbol\beta$ using:
\begin{equation}
\mathbf{R} \boldsymbol\beta = \boldsymbol\theta, \quad \text{where } \boldsymbol\theta = \mathbf{Q}^\mathsf{T} \mathbf{Y}
\label{eq:qr_beta}
\end{equation}
Since $\mathbf{R}$ is upper triangular, solving for $\boldsymbol\beta$ requires only back-substitution, which is numerically stable and avoids direct matrix inversion.

\subsubsection{Incremental updates}
A key advantage of QR decomposition is its ability to incrementally update the regression model when new observations arrive. Instead of re-computing the complete decomposition from scratch, the upper triangular matrix $\mathbf{R}$ can be efficiently updated by adding a new row. 

Following the approach in \cite{10.1093/imamat/12.3.329}, we construct an augmented matrix:
\begin{equation}
\mathbf{A}_\mathbf{XY} =
\begin{bmatrix}
\mathbf{X} & \mathbf{Y}
\end{bmatrix}
\end{equation}
Applying QR decomposition to $\mathbf{A}_\mathbf{XY}$ yields an upper triangular matrix $\mathbf{R_{A_{XY}}}$ of size $(p+1) \times (p+1)$:
\begin{equation}
\mathbf{R_{A_{XY}}} =
\begin{bmatrix}
r_{11} & r_{12} & \cdots & \theta_1 \\
0 & r_{22} & \cdots & \theta_2 \\
\vdots & \vdots & \vdots & \vdots \\
0 & \cdots & \cdots & \sqrt{RSS}
\end{bmatrix}
\label{eq:ra}
\end{equation}
where $\sqrt{RSS}$ represents the square root of the residual sum of squares.

To incorporate a new observation $(\mathbf{x}_\text{new}, y_\text{new}) \doteq (x_{11} \cdots x_{1p},  y_1)$, we append a new row to $\mathbf{R_{A_{XY}}}$:
\begin{equation}
\mathbf{R_{A_{XY}}} =
\begin{bmatrix}
r_{11} & r_{12} & \cdots & \theta_1 \\
0 & r_{22} & \cdots & \theta_2 \\
\vdots & \vdots & \vdots & \vdots \\
0 & \cdots & \cdots & \sqrt{RSS} \\
x_{11} & \cdots & x_{1p} & y_1
\end{bmatrix}
\label{eq:add_row}
\end{equation}
Applying QR decomposition to the updated matrix efficiently incorporates the new observation without recomputing from scratch.

\subsubsection{Distributed computation}
\label{sec:distLM}
Beyond streaming updates, QR decomposition also facilitates distributed computation, where different nodes process subsets of data and later combine the results. Each node $i$ computes its own local QR decomposition:
\begin{equation}
\mathbf{R}_{local}^i = \text{QR}(\mathbf{X}^i, \mathbf{Y}^i)
\end{equation}
and transmits only the resulting $\mathbf{R}_{local}^i$ matrices, rather than the full dataset. The global matrix:
\begin{equation}
\mathbf{R}_{G} =
\begin{bmatrix}
\mathbf{R}_{local}^1 \\
\mathbf{R}_{local}^2 \\
\vdots \\
\mathbf{R}_{local}^n
\end{bmatrix}
\label{eq:add_r}
\end{equation}
is then combined via another QR decomposition to produce the final upper triangular matrix, from which $\boldsymbol\beta$ is computed.

This approach reduces communication overhead and enables parallel processing; however, it raises privacy concerns, as each node must share at least some summary statistics of its data. A potential solution, to be discussed in Section \ref{sec:privacy}, is to use secure aggregation methods to compute the global model without exposing observations at node-level.

Previous studies have explored the use of QR decomposition in parallel computing environments. In particular, \cite{doi:10.1137/080731992} proposed a parallelized QR decomposition method, where a matrix $\mathbf{A}$ is horizontally partitioned into multiple slices:

\begin{equation}
\mathbf{A} = 
\begin{bmatrix}
\mathbf{A}_0 \\ \mathbf{A}_1 \\ \mathbf{A}_2 \\ \mathbf{A}_3 
\end{bmatrix}
=
\begin{bmatrix}
\mathbf{Q}_0 \mathbf{R}_0 \\ \mathbf{Q}_1 \mathbf{R}_1 \\ \mathbf{Q}_2 \mathbf{R}_2 \\ \mathbf{Q}_3 \mathbf{R}_3 
\end{bmatrix}
\end{equation}

where each slice undergoes independent QR decomposition. The intermediate $\mathbf{R}_i$ matrices are then iteratively combined:

\begin{equation}
\begin{bmatrix}
\begin{array}{c}
\mathbf{R}_0 \\ \mathbf{R}_1 \\ 
\hline
\mathbf{R}_2 \\ \mathbf{R}_3
\end{array}
\end{bmatrix}
=
\begin{bmatrix}
\begin{array}{c}
\begin{bmatrix}
\mathbf{R}_0 \\ \mathbf{R}_1
\end{bmatrix}
\\
%\hline
\begin{bmatrix}
\mathbf{R}_2 \\ \mathbf{R}_3
\end{bmatrix}
\end{array}
\end{bmatrix}
=
\begin{bmatrix}
\mathbf{Q}_{01} \mathbf{R}_{01} \\ \mathbf{Q}_{23} \mathbf{R}_{23}
\end{bmatrix}
\label{eq:slice_R}
\end{equation}

until a final global $\mathbf{R}$ matrix is obtained:

\begin{equation}
\begin{bmatrix}
\mathbf{R}_{01} \\ \mathbf{R}_{23}
\end{bmatrix}
=
\mathbf{Q}_{0123} \mathbf{R}_{0123}
\label{eq:demmel_final_update}
\end{equation}

While these methods focus on optimizing computational efficiency, they typically assume centralized control and uniform data distribution across nodes. In contrast, real-world scenarios often involve heterogeneous data sources and privacy constraints, which motivate the need for alternative approaches, such as the one proposed in this work.

\section{Privacy-preserving adaptations} 
\label{sec:privacy}

Privacy-preserving distributed algorithms let a network of participants compute a shared result while keeping each node's sensitive data confidential.

From \eqref{eq:add_r}, it is possible to introduce a privacy-preserving distributed algorithm that fits a linear regression model without having access to all the observations, only by passing the aggregated matrix $\mathbf{R}_{local}$ to other nodes.
In this case, privacy is for the original data in each node, not the shared model.
Data communication channels are assumed to be made private to the peers using standard cryptographic methods.

\begin{algorithm}
\caption{Distributed linear regression model}
\label{alg:dist_lm}
\begin{algorithmic}
\Require $\mathbf{X}_{local}, \mathbf{Y}_{local}, \mathbf{V}$
\State $\mathbf{A_{XY}} \gets [\mathbf{X}_{local} \quad \mathbf{Y}_{local} ]$
\State $(\mathbf{Q}, \mathbf{R}_{local}) \gets$ \texttt{decqr($\mathbf{A_{XY}}$)}
\Comment{QR decomposition}

\Comment{for loops can be done concurrently}
\For{$v \in \mathbf{V}$} \Comment{$\mathbf{V}$ is list of all nodes}
    \State \texttt{send($v$, $\mathbf{R}_{local}$)}
\EndFor
\For{$v \in \mathbf{V}$}
    \State $\mathbf{R}_{local} \gets 
    \begin{bmatrix}
    \mathbf{R}_{local} \\
    \texttt{receive($v$, $\mathbf{R}_{remote}$)}
    \end{bmatrix}
    $
\EndFor

\State $(\mathbf{Q}, \mathbf{R}_{global}) \gets$ \texttt{decqr($\mathbf{R}_{local}$)}
\State $\begin{bmatrix} \mathbf{R} \quad \quad \quad \boldsymbol\theta \\
\mathbf{0} \quad \sqrt{RSS}\end{bmatrix} \gets \mathbf{R}_{global}$
\State $\boldsymbol{\hat{\beta}} \gets$ \texttt{backsolve($\mathbf{R}$, $\mathbf{\boldsymbol\theta}$)}
\end{algorithmic}
\end{algorithm}

Instead of computing the QR decomposition in a distributed manner, for which not all methods are privacy-preserving in a federated setting \citep{10207753}, Algorithm \ref{alg:dist_lm} shows that it's possible to only share an aggregated matrix from local data and still arrive at the same result as in a centralized setting. Furthermore, in a setting where observations are produced in different places and then transmitted to a central location where the final computation takes place, Algorithm \ref{alg:dist_lm} reduces communication due to the size properties of the sent matrix, albeit at the expense of more computation.

Both Algorithm \ref{alg:dist_lm} and the follow up algorithm, that we will introduce in Section \ref{sec:glm}, assume a generous system model where: each node knows and can communicate with all other nodes; the network is reliable, messages are not corrupted, duplicated or lost and FIFO order is preserved in each channel; nodes do not crash and the membership is stable. The algorithms may have a cycle loop and proceed in rounds and have an initial computation phase, data transmission to all nodes, data reception from all nodes, and another computation phase. While it is possible that these algorithms can be adapted to provide message fault tolerance and support a less permissive system model, in this paper we focus on the initial introduction of these new approaches in a more simple setting. 

Privacy is preserved since although the QR decomposition of a matrix is unique \citep{10.5555/248979}, from only the shared matrix, $\mathbf{R}_{local}$, it is not possible to recover the original data, due to the fact that exists an infinite number of matrices $\mathbf{Q}$ that could be multiplied by $\mathbf{R}_{local}$ reaching different results. Proof outlined in Appendix \ref{sec:privacy_r}.

Compared to cryptographic techniques such as Secure Multiparty Computation (SMC) \citep{froelicher2021scalable} and Homomorphic Encryption (HE) \citep{Hall2011SecureML, gascon_privacy-preserving_2017}, which provide strong theoretical privacy guarantees but at a high computational cost, the proposed approach offers a lighter alternative by leveraging structural transformations instead of encryption.
Also, unlike Differential Privacy (DP), which ensures privacy by injecting calibrated noise to obscure individual contributions \citep{10.1007/11681878_14, 10.1145/1806689.1806787}, the presented method avoids utility loss by relying on the non-recoverability of the shared matrix $\mathbf{R}_{local}$.
While this results in weaker formal security guarantees compared to SMC and DP \citep{10.1145/2818000.2818027}, it provides a practical trade-off between privacy and computational efficiency.

\section{Extending to generalized linear regression}
\label{sec:glm}

Having presented the linear model in Section \ref{sec:lm}, which assumes that the residuals from linear regression follow a Gaussian distribution, we now explore a more generalized framework. In many applications, observations deviate from a Gaussian distribution, with binary outcomes or count data serving as prominent examples.

For such cases, different assumptions must be made, which is where the generalized linear model (GLM) \citep{doi:10.2307/2344614} comes into play. The GLM extends linear regression by assuming that the response variable follows a distribution from the exponential family. In this framework, the variance of observations is naturally linked to the mean, and the mean $\mu$ is related to a linear predictor $\eta$ through a nonlinear link function.

Although the GLM does not have closed-form estimation equations, it can still be expressed similarly to \eqref{eq:normal_lm} via its normal equations:
\begin{equation}
(\mathbf{X}^\mathsf{T}\mathbf{W}\mathbf{X}) \boldsymbol{\hat{\beta}} = \mathbf{X}^\mathsf{T} \mathbf{W} \mathbf{z}
\label{eq:normal_glm_with_z}
\end{equation}
and estimated using various iterative algorithms. Here, the diagonal matrix $\mathbf{W}$ and the transformed response vector $\mathbf{z}$ are defined as:

\begin{equation}
w_i = \frac{(\frac{\partial \mu_i}{\partial \eta_i})^2}{\mathrm{var}(\mathrm{Y}_i)} = \frac{ (\frac{\partial \mu_i}{\partial \eta_i})^2}{\alpha(\phi) V(\mu_i)}
\label{eq:glm_w}
\end{equation}

\begin{equation}
\mathbf{z} = \eta + \frac{\mathbf{Y} - \mu(\eta)}{\frac{\partial \mu_i}{\partial \eta_i}}
\label{eq:z}
\end{equation}
where $\alpha(\phi)$ is a dispersion parameter function with $\phi$ standing as a type of nuisance parameter such as the variance of a normal distribution or the shape parameter of a gamma distribution, and $V(\mu)$ is the variance function, both of which depend on the specific distribution of the observations.

One of the widely used estimation methods for GLMs is maximum likelihood estimation (MLE), which can be computed iteratively by solving the weighted least squares equation \eqref{eq:normal_glm_with_z} while updating the variances based on the model.

Similar to \eqref{eq:lm_beta_hat}, equation \eqref{eq:normal_glm_with_z} can be solved using multiple weighted least squares (WLS), which extends ordinary least squares to account for heteroscedasticity. The MLE of a GLM is therefore obtained using iteratively reweighted least squares (IRLS) \citep{00359246}, where the weights and response are updated iteratively until convergence:
\begin{equation}
\boldsymbol{\hat{\beta}} = (\mathbf{X}^\mathsf{T} \mathbf{W}\mathbf{X})^{-1} \mathbf{X}^\mathsf{T} \mathbf{W} \mathbf{z}
\label{eq:glm_beta}
\end{equation}

The IRLS approach has a known parallel-distributed implementation \citep{10.4310/SII.2013.v6.n4.a15}, where the main idea is to decompose the weighted least squares computation and distribute the tasks across multiple computing nodes:
\begin{equation}
\begin{aligned}
\mathbf{X}^\mathsf{T} \mathbf{W} \mathbf{X} & = \sum_{i=1}^{k} \mathbf{X}_i^\mathsf{T} \mathbf{W}_i \mathbf{X}_i, \\
\mathbf{X}^\mathsf{T} \mathbf{W} \mathbf{z} & = \sum{i=1}^{k} \mathbf{X}_i^\mathsf{T} \mathbf{W}_i \mathbf{z}_i
\end{aligned}
\label{eq:decomp_w}
\end{equation}
where $k$ is the number of computing nodes participating in the computation.

However, this method suffers from the numerical stability issues discussed in Section \ref{sec:lm}, as the often ill-conditioned nature of the cross-product matrix $\mathbf{X}^\mathsf{T} \mathbf{W}\mathbf{X}$ can lead to inaccuracies.

Alternatively, by returning to \eqref{eq:normal_glm_with_z} and considering that $\mathbf{W} \in \mathbb{R}^{n \times n}$ is a diagonal matrix, we can define transformed variables:

\begin{equation}
\mathbf{\tilde{X}} = \sqrt{\mathbf{W}}\mathbf{X}, \quad \mathbf{\tilde{z}} = \sqrt{\mathbf{W}}\mathbf{z}
\end{equation}
which allow us to rewrite \eqref{eq:normal_glm_with_z} as:

\begin{equation}
\mathbf{\tilde{X}}^\mathsf{T}\mathbf{\tilde{X}} \boldsymbol{\hat{\beta}} = \mathbf{\tilde{X}}^\mathsf{T} \mathbf{\tilde{z}}
\label{eq:tilde_xz}
\end{equation}

This equation can be solved iteratively using ordinary least squares, as described in Section \ref{sec:qr}. Algorithm \ref{alg:dist_glm} outlines this approach.

\begin{algorithm}
\caption{Distributed generalized linear model}
\label{alg:dist_glm}
\begin{algorithmic}
\Require $\mathbf{X}_{local}, \mathbf{Y}_{local}, \boldsymbol{\hat{\beta}}, \texttt{family}, \texttt{maxit}, \texttt{tol}, \mathbf{V}$
\State $i \gets 0$
\While{$i \leq \texttt{maxit}$}
\State $\boldsymbol\eta \gets \mathbf{X}_{local} \boldsymbol{\hat{\beta}}$
\State $\boldsymbol\mu \gets g^{-1}(\boldsymbol\eta)$ \Comment{inverse of link function $g(.)$ from \texttt{family}}
\State $\boldsymbol\mu' \gets \frac{\partial \mu}{\partial \eta}$
\State $\mathbf{z} \gets \boldsymbol\eta + \frac{\mathbf{Y}_{local} - \boldsymbol\mu}{\boldsymbol\mu'}$
\State $\mathbf{W} \gets \frac{\boldsymbol\mu'^2}{V(\boldsymbol\mu)}$
\Comment{$V(.)$ is the variance function from \texttt{family}}
\State $\mathbf{\tilde{X}}_{local} \gets \sqrt{\mathbf{W}}\mathbf{X}_{local}$
\State $\mathbf{\tilde{z}} \gets \sqrt{\mathbf{W}}\mathbf{z}$
\State $\mathbf{A_{XY}} \gets [\mathbf{\tilde{X}}_{local} \quad \mathbf{\tilde{z}} ]$
\State $(\mathbf{Q}, \mathbf{R}_{local}) \gets$ \texttt{decqr($\mathbf{A_{XY}}$)}
\Comment{QR decomposition}

\Comment{for loops can be done concurrently}
\For{$v \in \mathbf{V}$} \Comment{$\mathbf{V}$ is list of all nodes}
    \State \texttt{send($v$, $\mathbf{R}_{local}$)}
\EndFor
\For{$v \in \mathbf{V}$}
    \State $\mathbf{R}_{local} \gets 
    \begin{bmatrix}
    \mathbf{R}_{local} \\
    \texttt{receive($v$, $\mathbf{R}_{remote}$)}
    \end{bmatrix}
    $
\EndFor
\State $(\mathbf{Q}, \mathbf{R}_{global}) \gets$ \texttt{decqr($\mathbf{R}_{local}$)}
\State $\begin{bmatrix} \mathbf{R} \quad \quad \quad \boldsymbol\theta \\
\mathbf{0} \quad \sqrt{RSS}\end{bmatrix} \gets \mathbf{R}_{global}$
\State $\boldsymbol{\hat{\beta}}_{old} \gets \boldsymbol{\hat{\beta}}$
\State $\boldsymbol{\hat{\beta}} \gets$ \texttt{backsolve($\mathbf{R}$, $\mathbf{\boldsymbol\theta}$)}
\If{$\max \vert \frac{\boldsymbol{\hat{\beta}}_{old}-\boldsymbol{\hat{\beta}}}{\sqrt{\texttt{vcov($\mathbf{R}_{global}$)}}} \vert < \texttt{tol}$}
\Comment{\texttt{vcov} returns the diagonal of the}
\State \texttt{break}
\Comment{variance-covariance matrix}
\EndIf
\State $i \gets i + 1$
\EndWhile
\end{algorithmic}
\end{algorithm}

Although various QR decomposition methods exist for \eqref{eq:qr}, not all are suitable for $\mathbf{\tilde{X}}$, as both $\mathbf{X}$ and $\mathbf{W}$ may be rank-deficient, leading to instability \citep{10.1093/oso/9780198535645.003.0006, ANDERSON1992243}.

Considering the limitations of floating-point arithmetic \citep{8766229}, implementing an algorithm that monitors $\sqrt{\mathbf{W}}$ within an acceptable tolerance is crucial for ensuring numerical stability.\\

\section{Experimental evaluation}
\label{sec:experimental_eval}

We now present application examples to assess and illustrate the practical usefulness of the methodology proposed in this paper for both linear regression and generalized linear regression. First, these methods are applied to simulated data, followed by examples using real datasets.

\subsection{Numerical studies with simulated data}

Both approaches, the linear model and generalized linear model algorithms, have been implemented and can be referenced \footnote{\url{https://github.com/dbtnc/distributed_glm}}.
In this section, we perform a set of experiments to evaluate the approaches in terms of accuracy and compare them to the state-of-the-art centralized implementation. All experiments were run in a system with 2 CPUs and 8 GB RAM, running a simulation of each target distributed system setup.

\subsubsection{Distributed linear regression}
\label{subsec:exp_lm}

The initial step involved comparing a centralized version of Algorithm \ref{alg:dist_lm} with the standard R implementation (\texttt{lm}) from the included \texttt{stats} package \citep{Rcitation}. The objective was to establish a baseline and validate the implementation's correctness.
The comparison was conducted across various data sizes (with the number of observations set to 100, 1000 or 10000 and the number of predictors to 1, 3, 5, or 10). We estimated the $\boldsymbol{\hat{\beta}}$ regression coefficients and assessed differences, if any, from the standard R implementation. Our metric of choice was the Mean Absolute Error (MAE) across all regression coefficients, with 100 replicas for each combination, in order to minimize discrepancies between the pseudo-randomly generated datasets. In each replication, the generated dataset assumed $\boldsymbol\beta$ value(s) of 3, with an added error perturbation $\boldsymbol\varepsilon$ following a Gaussian distribution with mean 0 and standard deviation 1.

When evaluating the centralized version in comparison to the R implementation, null differences, in the MAE, were observed up to the limits of current floating-point arithmetic, leading us to conclude that the proposed modified OLS implementation produces consistent results.

To assess the distributed version described in Section \ref{sec:distLM}, we employed a shared memory approach, by splitting the original dataset in equally sized pieces across the simulated nodes.
For each dataset split piece, a single iteration of Algorithm \ref{alg:dist_lm} was performed, in order to simulate the behaviour of a single node.
Since the computation is distributed among the nodes and the accumulated matrix $\mathbf{R}_{global}$ can receive other matrices $\mathbf{R}_{remote}$ in any order, results may vary between nodes and differ from the centralized version.
For practical purposes, we consider this method similar to the centralized one, as the highest difference observed was on the order of $1 \times 10^{-15}$ for 5 simulated nodes (see Table \ref{tab:res_lm}).
It is worth noting that, as anticipated, the errors increase slightly with the number of predictors.

\begin{table}[ht]
\caption{Distributed linear regression with 5 simulated nodes: mean absolute differences for 100 replicas, with varying number of observations and predictors.}
\centering
\begin{tabular}{lcccc}
\toprule
Observations & \multicolumn{4}{c}{Predictors} \\
\cmidrule(lr){2-5}
             & 1           & 3           & 5           & 10          \\
\midrule
100    & 6.904e-16 & 1.083e-15 & 1.177e-15 & 1.445e-15 \\
1000   & 1.657e-15 & 2.418e-15 & 2.629e-15 & 2.912e-15 \\
10000  & 4.768e-15 & 7.954e-15 & 8.954e-15 & 8.878e-15 \\
\bottomrule
\end{tabular}
\label{tab:res_lm}
\end{table}

\begin{figure}[t]
\centering
\includegraphics[width=\textwidth]{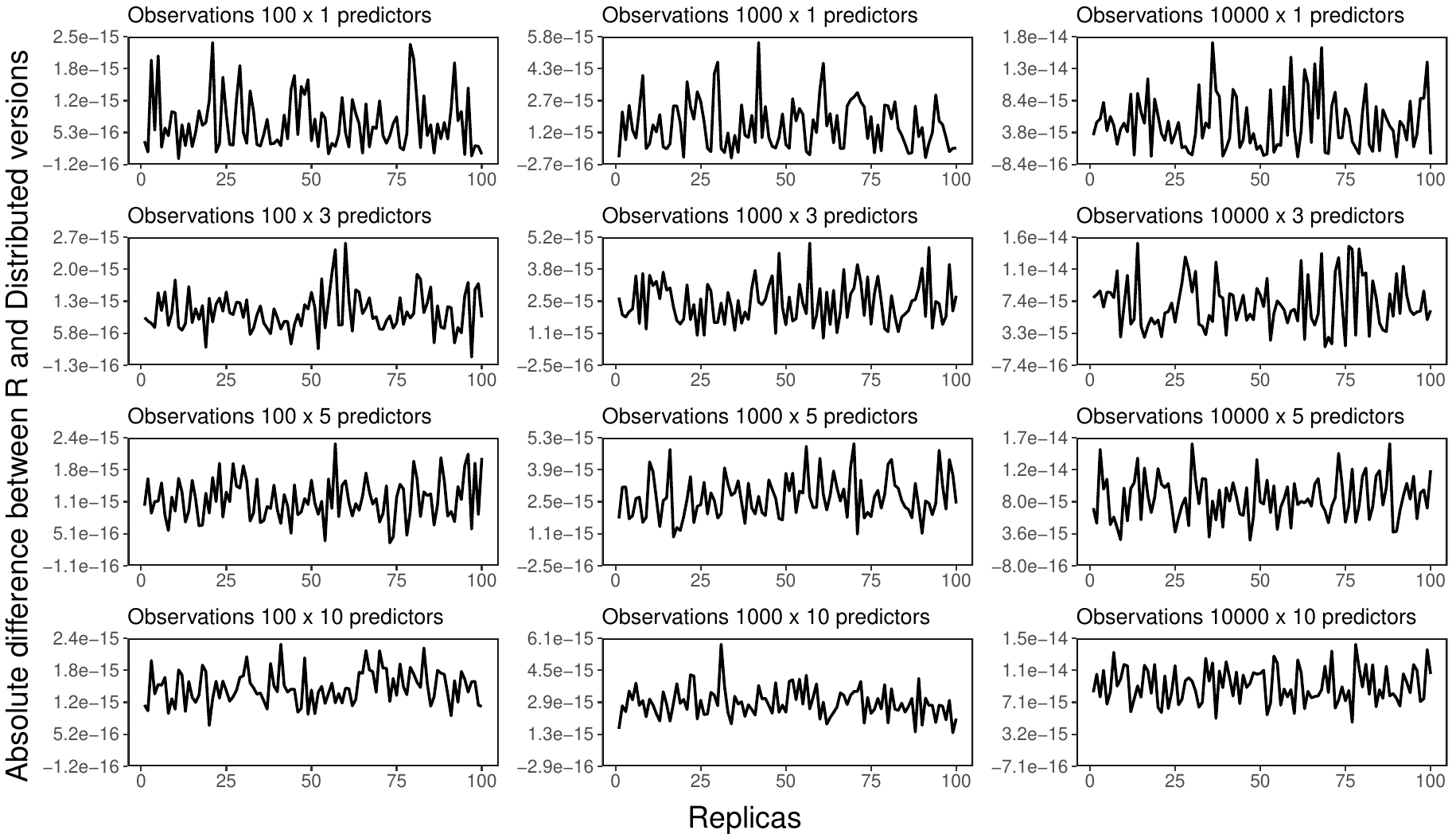}
\caption{Absolute difference between R \texttt{lm} function and LM distributed algorithm version with the number of observations set to 100, 1000 or 10000 and the number of predictors to 1, 3, 5, or 10 along 100 replicas.}
\label{fig:ae_lm}
\end{figure}

Figure \ref{fig:ae_lm} presents the results for the absolute difference between R \texttt{lm} function and our distributed algorithm with the number of observations set to 100, 1000 or 10000 and the number of predictors to 1, 3, 5, or 10 for 100 replicas with 5 simulated nodes.
Results show that besides the absolute difference between the centralized and distributed versions, there also is not a significant difference between these 100 replicas.

\subsubsection{Distributed generalized linear regression}

We now focus on evaluating the distributed approach for the generalized linear model. To assess its performance, we implemented both a centralized and a high-level distributed version to compare their results with the standard R implementation (\texttt{glm}) from the \texttt{stats} package. The objective was to empirically confirm whether the produced results were identical or sufficiently similar.

For the generalized regression case, the dataset was generated similarly to that of the linear model. The key distinction is that the response variable was not generated through a linear relationship but rather through a probabilistic relationship following the logistic distribution, also known as the logit function. This modification illustrates the flexibility of the generalized case, demonstrating its applicability to distributions beyond the Gaussian distribution. As a result, the response variable follows a Binomial distribution, with probabilities derived from the link function represented by the logistic distribution.

\begin{figure}[t]
\centering
\includegraphics[width=\textwidth]{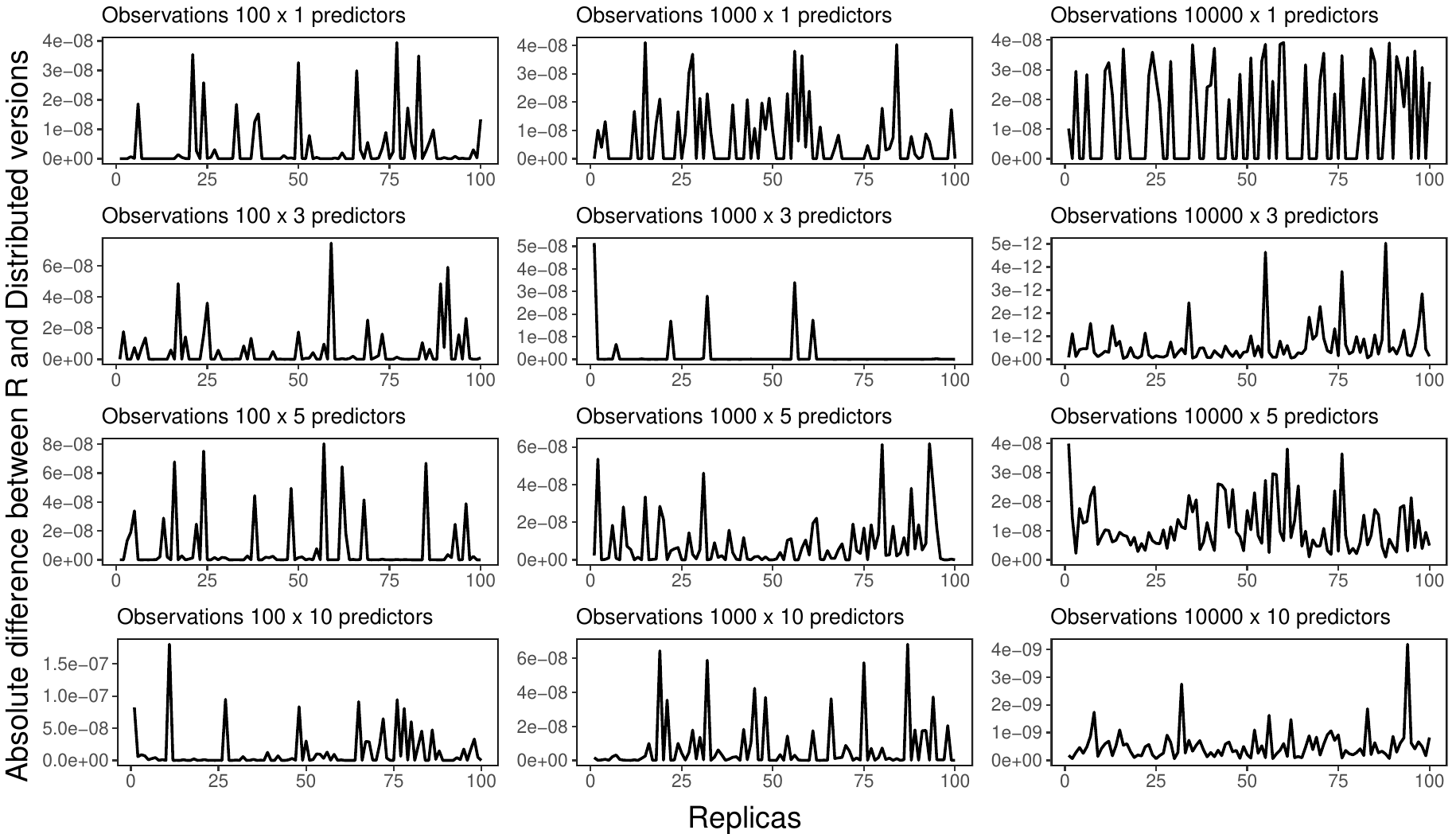}
\caption{Absolute difference between R \texttt{glm} function and GLM Distributed algorithm version with the number of observations set to 100, 1000 or 10000 and the number of predictors to 1, 3, 5, or 10 along 100 replicas.}
\label{fig:ae_glm}
\end{figure}

As observed in the linear model, the methods perform equivalently to the R implementation for practical purposes, with observed differences in mean absolute errors on the order of $10^{-8}$ (see Table \ref{tab:res_glm}).
Figure \ref{fig:ae_glm} presents the results for the absolute differences between the R \texttt{glm} function and the distributed generalized regression. This leads us to conclude that the methods perform similarly in practical terms.

\begin{table}[ht]
\caption{Distributed generalized linear regression with 5 simulated nodes: mean absolute errors for 100 replicas, with varying numbers of observations and predictors.}
\centering
\begin{tabular}{lcccc}
\toprule
Observations & \multicolumn{4}{c}{Predictors} \\
\cmidrule(lr){2-5}
             & 1           & 3           & 5           & 10          \\
\midrule
100    & 4.348e-09 & 7.985e-09 & 5.750e-09 & 4.907e-09 \\
1000   & 5.886e-09 & 1.051e-09 & 1.503e-08 & 6.895e-09 \\
10000  & 1.267e-08 & 6.814e-13 & 1.145e-08 & 5.645e-10 \\
\bottomrule
\end{tabular}
\label{tab:res_glm}
\end{table}

In addition to the configurations previously described and presented in Tables \ref{tab:res_lm} and \ref{tab:res_glm}, a similar experiment was conducted to further assess the horizontal scalability of the distributed algorithm for both the LM and GLM settings. This experiment used 100,000 observations, with the number of virtual nodes ranging from 10 to 100 in increments of 5, while keeping the number of predictors at 1, 3, or 5. Both the number of replicas (100) and the value of $\boldsymbol{\beta}$ (3) remained unchanged.
The results are consistent with those of the smaller previous experiment and can be found in Section \ref{sec:n_vnodes} of the Appendix. 
In the future, more complex experiments can be conducted under asymmetric data distributions across nodes, ideally using real-world datasets.

Both experiments, in the LM and GLM settings, lead to the conclusion that the distributed execution and associated data summarization—which enforce privacy preservation at each node—did not impact accuracy compared to classic centralized implementations. This is a fortunate outcome and a strong argument in favour of the practical deployment of this class of algorithms.\\

\subsection{Datasets \--- Diamonds and Credit Cards}

Having shown the results for synthetic data, our objective is to compare centralized approaches -- specifically, linear regression and generalized linear regression -- with their distributed counterparts, in a real-world application.
For the distributed implementation, we utilized Elixir \citep{elixir} and BEAM \citep{erlang} technologies to simulate a multi-machine computing environment, where each runtime process is responsible for a subset of the original dataset.
In turn, these processes communicate with each other through message passing, achieving the same effect as multiple connect machines.

To demonstrate flexibility in data distribution among participants, we set up seven nodes, though this number can be adjusted as long as it aligns with the algorithm's constraints. This setup ensures that each node handles an equal share of observations, except for one node, which manages any remaining data.

To extend our evaluation from synthetic data to real-world applications, we conducted experiments using the diamonds \footnote{https://ggplot2.tidyverse.org/reference/diamonds.html} and CreditCard \citep{Greene2003Econometric} datasets, for the LM and GLM setting respectively.

\begin{figure}[t]
\centering
\includegraphics[width=0.6\textwidth]{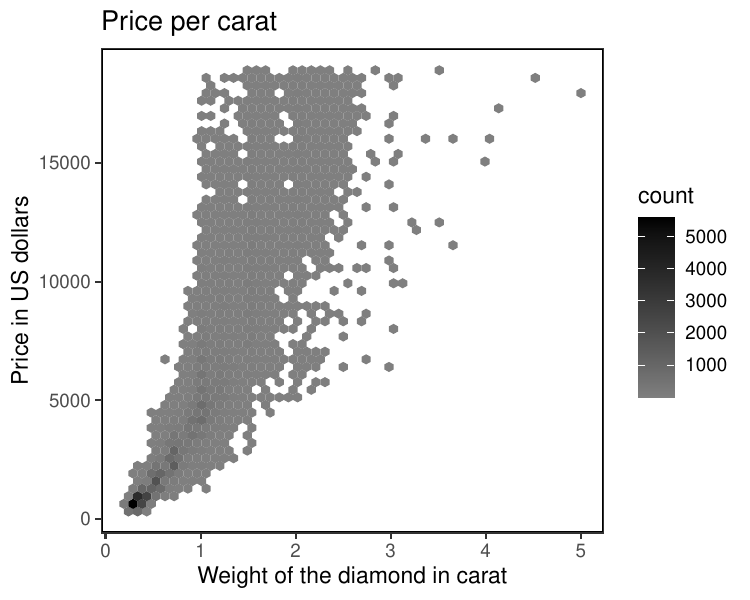}
\caption{Scatterplot of the relation of price and carat in a hexagon bin count format, with the number of bins set to 50.}
\label{fig:diamonds_carat}
\end{figure}

The diamonds dataset, which contains several attributes related to diamonds, has 53,940 rows and 10 variables, of which we are interested in predicting the quantitative variable price, set in US dollars, using other available variables.
Upon statistical analysis, the variables chosen as predictors were the quantitative variable carat (see Figure \ref{fig:diamonds_carat}) and the qualitative variables clarity --- measurement of diamond clarity, from I1 (worst) to IF (best) --- and color --- colour of the diamond, from D (best) to J (worst) --- (see Figure \ref{fig:diamonds_facet} in Appendix).

\begin{table}[ht]
\caption{Coefficients estimates for the diamonds dataset shared by centralized and distributed approaches.}
\centering
\begin{tabular}{lrlrlr}
\toprule
\multicolumn{2}{c}{Numeric} & \multicolumn{4}{c}{Categorical} \\
\cmidrule(lr){3-6}
             &      & clarity    &       & color    &     \\
\midrule
(Intercept) & -6699.9456 & clarity.SI2 & 2832.6514 & color.E & -216.4463\\
carat & 8856.2307 & clarity.SI1 & 3795.4712 & color.F & -314.9199\\

& & clarity.VS2 & 4466.1030 & color.G & -509.0893\\
& & clarity.VS1 & 4785.7910 & color.H & -985.0061\\
& & clarity.VVS2 & 5234.1629 & color.I & -1441.7666\\
& & clarity.VVS1 & 5351.8484 & color.J & -2340.8253\\
& & clarity.IF & 5718.2294\\

\bottomrule
\end{tabular}
\label{tab:diamonds_lm}
\end{table}

Both the centralized R \texttt{lm} version and the distributed version presented in Algorithm \ref{alg:dist_lm} achieve the same results up to floating point precision, which are presented in Table \ref{tab:diamonds_lm}. From these coefficients estimates, it is possible to conclude that, for example, for every extra 1 carat increased to the weight of the diamond, the expected price should increase by 8,856.23 US dollars.

Moving to the CreditCard dataset, a collection of data on a sample of individual applicants, for a type of credit card, regarding their credit card usage history and other personal and financial aspects.
With 1,319 rows and 12 variables, the variable of interest will be card, which represents if the application for a credit card was accepted.
Being this variable binary, the linear model assuming the Gaussian distribution for the residuals will not be adequate.
Instead, a generalized linear model with a Binomial distribution and an appropriate link function, such as the logit function, ensures that the predicted values remain within the valid range (0 to 1) for the probability of a credit card application being accepted. 
This approach accurately captures the relationship between the predictors and the binary outcome.

\begin{figure}[t]
\centering
\includegraphics[width=0.6\textwidth]{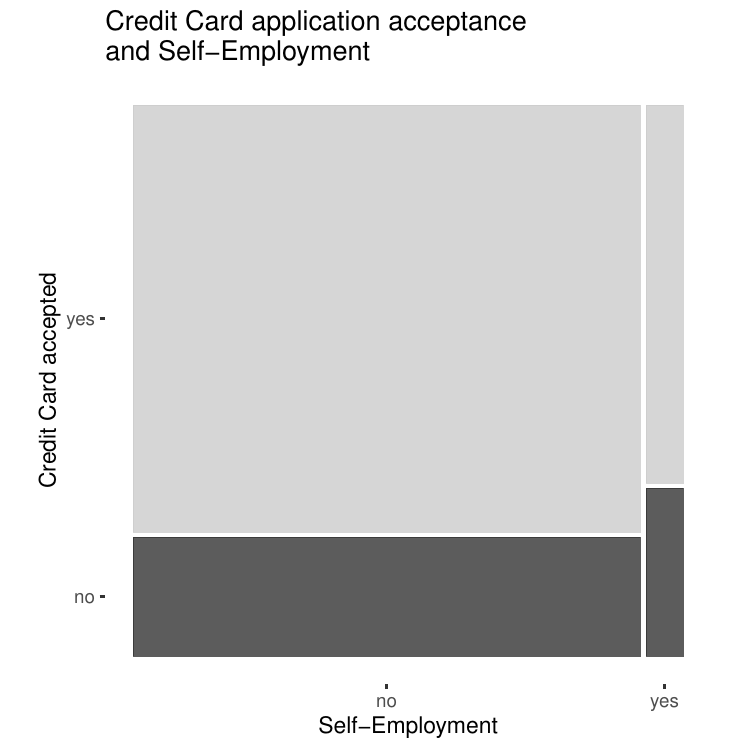}
\caption{Mosaic plot of the relation between credit card application acceptance and self-employment.}
\label{fig:cc_selfemp_mosaic}
\end{figure}

In order to demonstrate that the distributed algorithms support different types of data, in the same way that R \texttt{glm} does --- by converting non-numerical data to numerical representation in both the design matrix $\mathbf{X}$ and response matrix $\mathbf{Y}$ --- the chosen variables were the binary variable selfemp (see Figure \ref{fig:cc_selfemp_mosaic}), that represents if the individual is self-employed and the quantitative variable income, that represents the individual's yearly income in US dollars divided by 10,000.

Based on the coefficients estimates presented in Table \ref{tab:cc_glm}, it can be concluded that for a \$10,000 increase in yearly income, the odds of credit card application acceptance increase by a factor of 1.1835 (or 18.35\%), assuming the remaining predictor is held constant.
%IMPORTANTE: 1.1835=exp(0.16851)

\begin{table}[ht]
\caption{Coefficients estimates for the credit card dataset  shared by centralized and distributed approaches.}
\centering
\begin{tabular}{lrlr}
\toprule
\multicolumn{2}{c}{Numeric} & \multicolumn{2}{c}{Categorical} \\
\cmidrule(lr){3-4}
             &      & selfemp    &          \\
\midrule
(Intercept) & 0.73870 & selfemp.yes & -0.58753 \\
income & 0.16851 &  &  \\

\bottomrule
\end{tabular}
\label{tab:cc_glm}
\end{table}

To assess the similarity of results between centralized and distributed approaches, we applied the following pairwise comparison equation:
\begin{equation}
\lvert \boldsymbol{\hat{\beta}}_{distributed} - \boldsymbol{\hat{\beta}}_{central} \rvert \leq 10^{-8} + 10^{-5} * \lvert \boldsymbol{\hat{\beta}}_{central} \rvert
\end{equation}
for each process. Our findings showed that both approaches produced nearly identical coefficient values for both models, thereby demonstrating the effectiveness of these algorithms in real-world settings.

\section{Conclusion}

The paper presents two advancements in the field of linear regression models, with a focus on their application in distributed settings with data privacy concerns. The first contribution is the enhancement of linear models for distributed computation while ensuring data privacy. This is especially relevant for distributed and federated learning scenarios, where data is decentralized and privacy is paramount. The paper addresses this by introducing methods that allow linear models to be fitted in a distributed manner without compromising data privacy. These methods involve the use of QR decomposition and a privacy-preserving distributed algorithm that allows for the aggregation of data from multiple sources without requiring access to individual observations.

The second key contribution lies in the extension of these methodologies to Generalized Linear Models (GLMs). GLMs are more flexible than traditional linear models as they do not assume a normal distribution of error terms and can model a response variable as a nonlinear function of the predictors. The paper's approach to GLM computation in a distributed setting involves a novel algorithm that iteratively applies weighted least squares. This method accounts for the variance in observations and allows for the effective estimation of GLMs even with varying data distributions. 

Both methods were observed to be identical in practice to centralized approaches, since the expected variability in the data to be fitted largely exceeds any observable difference among the methods.

Additionally, this work offers a computationally efficient alternative to traditional privacy-preserving methods such as Secure Multiparty Computation (SMC), Homomorphic Encryption (HE), and Differential Privacy (DP).
The proposed approach ensures privacy through structural data transformations rather than cryptographic means or noise injection.
This avoids the computational overhead of encryption-based methods and the utility loss associated with DP, making it particularly well suited for streaming and distributed settings.
While this approach provides weaker formal privacy guarantees, it remains practical for semi-honest threat models where efficiency is a priority.

\paragraph{\textnormal{\textbf{Acknowledgements}}}
The first and second authors received support from the Portuguese Foundation for Science and Technology (FCT), provided through UID/00013: Centro de Matemática da Universidade do Minho (CMAT/UM).
The first author also received additional support from FCT through the Individual PhD Scholarship 2024.06508.BDANA.
The third author received support from national funds through FCT, under the reference UIDB/50014/2020.

\clearpage
\bibliography{\bib}
\clearpage
\appendix

% Enter appendix text:
\section{Privacy of sharing R}
\label{sec:privacy_r}

\begin{proposition}
It is not possible to recover matrix $\mathbf{X}$ from matrix $\mathbf{R}$ alone in the QR decomposition.
\end{proposition}

\begin{proof}
Given a matrix $\mathbf{X}$, its QR decomposition is:

\begin{equation}
\mathbf{X} = \mathbf{Q}\mathbf{R}
\end{equation}

where:
\begin{itemize}
    \item $\mathbf{Q}$ is an orthogonal matrix (i.e., $\mathbf{Q}^\mathsf{T}\mathbf{Q} = \mathbf{I}$),
    \item $\mathbf{R}$ is an upper triangular matrix.
\end{itemize}

The question is whether we can recover $\mathbf{X}$ if we only know $\mathbf{R}$, which by contradiction we will prove that it is not possible to recover $\mathbf{X}$ from $\mathbf{R}$ alone by showing that multiple different matrices $\mathbf{X}$ can have the same $\mathbf{R}$ in their QR decompositions.

\textit{Step 1:} Existence of Multiple Possible $\mathbf{Q}$-Matrices

Let $\mathbf{X_1}$ and $\mathbf{X_2}$ be two different matrices, each having a QR decomposition:

\begin{equation}
\mathbf{X_1} = \mathbf{Q_1} \mathbf{R_1} \quad \text{and} \quad \mathbf{X_2} = \mathbf{Q_2} \mathbf{R_2}
\end{equation}

If $\mathbf{X_1}$ and $\mathbf{X_2}$ have the same upper triangular matrix $\mathbf{R}$ (i.e., $\mathbf{R_1} = \mathbf{R_2} = \mathbf{R}$), this means:

\begin{equation}
\mathbf{X_1} = \mathbf{Q_1} \mathbf{R} \quad \text{and} \quad \mathbf{X_2} = \mathbf{Q_2} \mathbf{R}
\end{equation}

Therefore, the matrices $\mathbf{X_1}$ and $\mathbf{X_2}$ are related to $\mathbf{Q_1}$ and $\mathbf{Q_2}$, respectively. The matrix $\mathbf{R}$ is the same for both.

\textit{Step 2:} Construct Different $\mathbf{X_1}$ and $\mathbf{X_2}$

Now, we construct two different orthogonal matrices $\mathbf{Q_1}$ and $\mathbf{Q_2}$. For simplicity, assume $\mathbf{X_1}$ and $\mathbf{X_2}$ are $2 \times 2$ matrices, and let $\mathbf{Q_1}$ and $\mathbf{Q_2}$ be orthogonal matrices. For example:

\begin{equation}
\mathbf{Q_1} = \begin{bmatrix} 1 & 0 \\ 0 & 1 \end{bmatrix}, \quad \mathbf{Q_2} = \begin{bmatrix} 0 & -1 \\ 1 & 0 \end{bmatrix}
\end{equation}

Both $\mathbf{Q_1}$ and $\mathbf{Q_2}$ are orthogonal (since $\mathbf{Q_1}^\mathsf{T}\mathbf{Q_1} = \mathbf{I}$ and $\mathbf{Q_2}^\mathsf{T}\mathbf{Q_2} = \mathbf{I}$).

Let the upper triangular matrix $\mathbf{R}$ be:

\begin{equation}
\mathbf{R} = \begin{bmatrix} 1 & 2 \\ 0 & 1 \end{bmatrix}
\end{equation}

Now, compute $\mathbf{X_1}$ and $\mathbf{X_2}$:

\begin{equation}
\begin{gathered}
\mathbf{X_1} = \mathbf{Q_1} \mathbf{R} = \begin{bmatrix} 1 & 0 \\ 0 & 1 \end{bmatrix} \begin{bmatrix} 1 & 2 \\ 0 & 1 \end{bmatrix} = \begin{bmatrix} 1 & 2 \\ 0 & 1 \end{bmatrix} \\
\mathbf{X_2} = \mathbf{Q_2} \mathbf{R} = \begin{bmatrix} 0 & -1 \\ 1 & 0 \end{bmatrix} \begin{bmatrix} 1 & 2 \\ 0 & 1 \end{bmatrix} = \begin{bmatrix} 0 & -1 \\ 1 & 2 \end{bmatrix}
\end{gathered}
\end{equation}

Thus, we have two different matrices:

\begin{equation}
\mathbf{X_1} = \begin{bmatrix} 1 & 2 \\ 0 & 1 \end{bmatrix}, \quad \mathbf{X_2} = \begin{bmatrix} 0 & -1 \\ 1 & 2 \end{bmatrix}
\end{equation}

\textit{Step 3:} Conclusion

Both $\mathbf{X_1}$ and $\mathbf{X_2}$ have the same upper triangular matrix $\mathbf{R}$, but different orthogonal matrices $\mathbf{Q_1}$ and $\mathbf{Q_2}$. Therefore, even though they share the same $\mathbf{R}$, the matrices $\mathbf{X_1}$ and $\mathbf{X_2}$ are different.

This demonstrates that it's not possible to recover $\mathbf{X}$ from $\mathbf{R}$ alone because different matrices $\mathbf{X}$ can produce the same $\mathbf{R}$ in their QR decompositions. Therefore, both $\mathbf{Q}$ and $\mathbf{R}$ are necessary to recover $\mathbf{X}$.

\end{proof}

\section{N Virtual Nodes}
\label{sec:n_vnodes}

In order to test the horizontal scalability of the proposed Algorithms \ref{alg:dist_lm} (LM) and \ref{alg:dist_glm} (GLM), another experiment was conducted, where the generated datasets, followed the same settings used in Section \ref{sec:experimental_eval}, but with configurations of 10,000 and 100,000 rows, with 1, 3 and 5 predictors.
Instead of using the same number of nodes, the value of nodes varied from 10 to 100, with increments of 5.

\begin{figure}[t]
\centering
\includegraphics[width=\textwidth]{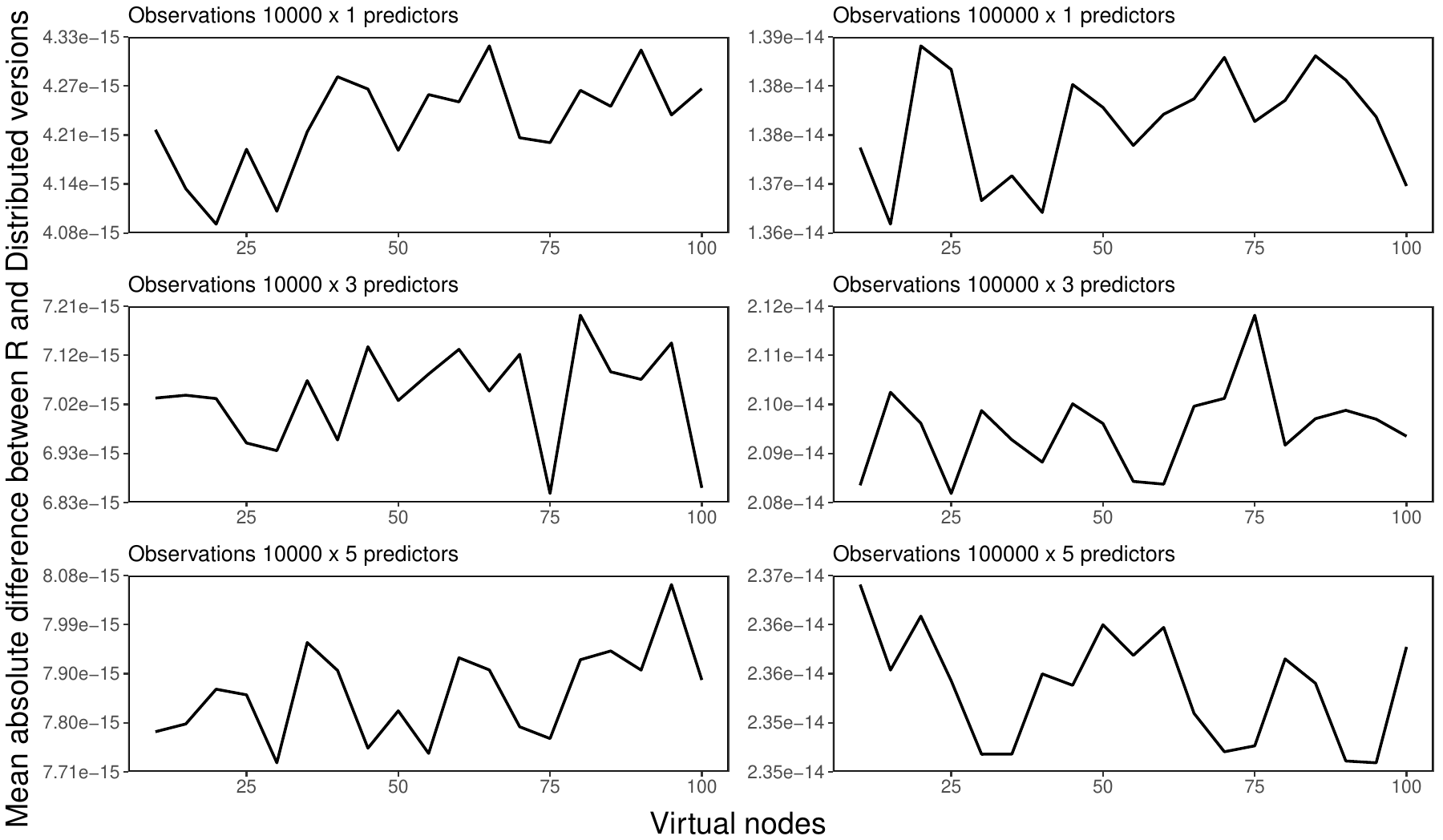}
\caption{Mean absolute difference between R \texttt{lm} function and LM Distributed algorithm version with the number of observations set to 10000 or 100000 and the number of predictors to 1, 3, or 5 along 100 replicas for the number of virtual nodes from 10 to 100 in increments of 5.}
\label{fig:mae_vnodes_lm}
\end{figure}

\begin{figure}[t]
\centering
\includegraphics[width=\textwidth]{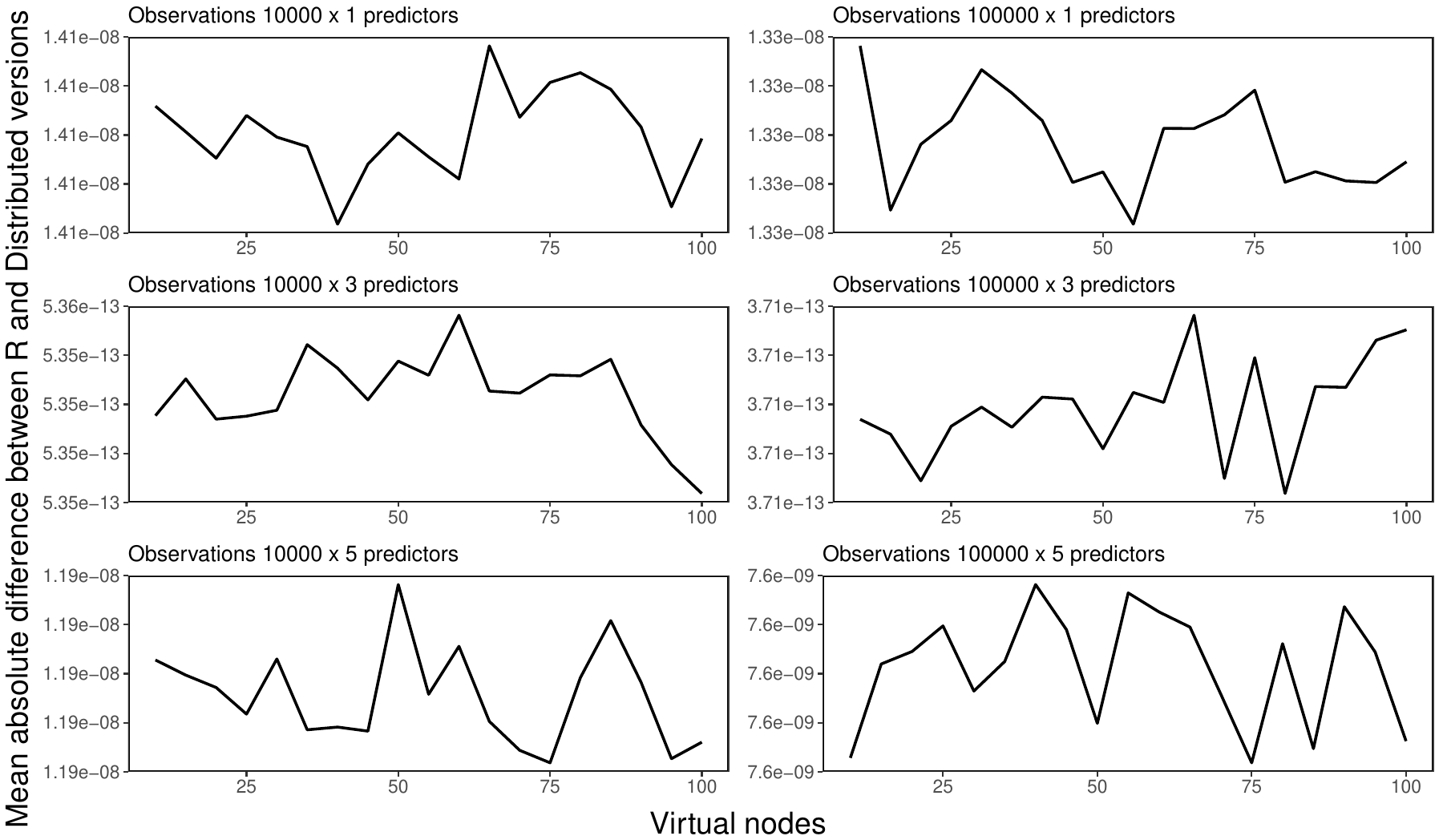}
\caption{Mean absolute difference between R \texttt{glm} function and GLM Distributed algorithm version with the number of observations set to 10000 or 100000 and the number of predictors to 1, 3, or 5 along 100 replicas for the number of virtual nodes from 10 to 100 in increments of 5}
\label{fig:mae_vnodes_glm}
\end{figure}

Table \ref{tab:res_vnodes_lm} shows the MAE for the different configurations for Algorithm \ref{alg:dist_lm}, and Table \ref{tab:res_vnodes_glm} shows the MAE for Algorithm \ref{alg:dist_glm}.
Figures \ref{fig:mae_vnodes_lm} and \ref{fig:mae_vnodes_glm}, demonstrate that the results do not vary significantly due to the number of nodes, for Algorithms \ref{alg:dist_lm} (LM) and \ref{alg:dist_glm} (GLM) respectively.

\begin{table}[ht]
\caption{Linear model: mean absolute errors for different configurations of observations and predictors for N simulated nodes.}
\begin{tabular}{lrrrrrr}
\toprule
& \multicolumn{6}{c}{Size} \\
\cmidrule(lr){2-7}
\# Nodes & $10000 \times 1$ & $10000 \times 3$ & $10000 \times 5$ & $100000 \times 1$ & $100000 \times 3$ & $100000 \times 5$ \\
%\midrule
\cmidrule(lr){1-1}
\cmidrule(lr){2-7}
10 & 4.213e-15 & 7.034e-15 & 7.787e-15 & 1.375e-14 & 2.086e-14 & 2.368e-14\\

15 & 5.541e-15 & 7.057e-15 & 8.369e-15 & 1.370e-14 & 2.348e-14 & 2.435e-14\\

20 & 5.251e-15 & 7.378e-15 & 8.493e-15 & 1.408e-14 & 2.066e-14 & 2.354e-14\\

25 & 4.808e-15 & 7.025e-15 & 8.338e-15 & 1.438e-14 & 2.251e-14 & 2.411e-14\\

30 & 5.196e-15 & 7.825e-15 & 8.240e-15 & 1.628e-14 & 2.216e-14 & 2.318e-14\\

35 & 4.904e-15 & 7.366e-15 & 8.219e-15 & 1.444e-14 & 2.060e-14 & 2.286e-14\\

40 & 5.507e-15 & 7.421e-15 & 8.559e-15 & 1.375e-14 & 2.006e-14 & 2.507e-14\\

45 & 5.097e-15 & 7.061e-15 & 8.195e-15 & 1.444e-14 & 2.117e-14 & 2.383e-14\\

50 & 5.120e-15 & 7.968e-15 & 8.395e-15 & 1.353e-14 & 2.199e-14 & 2.361e-14\\

55 & 5.260e-15 & 7.309e-15 & 8.724e-15 & 1.462e-14 & 2.103e-14 & 2.338e-14\\

60 & 4.805e-15 & 6.959e-15 & 8.829e-15 & 1.710e-14 & 2.054e-14 & 2.411e-14\\

65 & 4.443e-15 & 7.582e-15 & 8.274e-15 & 1.363e-14 & 2.205e-14 & 2.408e-14\\

70 & 5.535e-15 & 7.542e-15 & 8.269e-15 & 1.505e-14 & 2.255e-14 & 2.292e-14\\

75 & 5.261e-15 & 7.902e-15 & 8.067e-15 & 1.465e-14 & 2.311e-14 & 2.401e-14\\

80 & 4.566e-15 & 7.546e-15 & 8.559e-15 & 1.527e-14 & 2.207e-14 & 2.524e-14\\

85 & 4.735e-15 & 8.106e-15 & 8.468e-15 & 1.528e-14 & 2.463e-14 & 2.313e-14\\

90 & 5.464e-15 & 7.703e-15 & 8.512e-15 & 1.440e-14 & 2.220e-14 & 2.296e-14\\

95 & 5.327e-15 & 7.512e-15 & 8.412e-15 & 1.316e-14 & 2.372e-14 & 2.181e-14\\

100 & 5.121e-15 & 7.481e-15 & 8.011e-15 & 1.377e-14 & 2.019e-14 & 2.462e-14\\
\bottomrule
\end{tabular}
\label{tab:res_vnodes_lm}
\end{table}

\begin{table}[ht]
\caption{Generalized Linear model: mean absolute errors for different configurations of observations and predictors for N simulated nodes.}
\begin{tabular}{lrrrrrr}
\toprule
& \multicolumn{6}{c}{Size} \\
\cmidrule(lr){2-7}
\# Nodes & $10000 \times 1$ & $10000 \times 3$ & $10000 \times 5$ & $100000 \times 1$ & $100000 \times 3$ & $100000 \times 5$ \\
%\midrule
\cmidrule(lr){1-1}
\cmidrule(lr){2-7}
10 & 1.359e-08 & 1.206e-12 & 1.049e-08 & 1.857e-08 & 3.647e-13 & 7.811e-09\\

15 & 1.339e-08 & 8.262e-13 & 9.577e-09 & 1.413e-08 & 3.730e-13 & 7.849e-09\\

20 & 1.523e-08 & 5.609e-13 & 1.058e-08 & 1.647e-08 & 3.525e-13 & 7.741e-09\\

25 & 1.303e-08 & 5.856e-13 & 1.199e-08 & 1.517e-08 & 3.575e-13 & 7.771e-09\\

30 & 1.261e-08 & 7.410e-13 & 1.118e-08 & 1.816e-08 & 4.079e-13 & 7.526e-09\\

35 & 1.390e-08 & 5.401e-13 & 8.865e-09 & 1.341e-08 & 3.429e-13 & 7.992e-09\\

40 & 1.289e-08 & 6.728e-13 & 8.771e-09 & 1.454e-08 & 3.745e-13 & 8.021e-09\\

45 & 1.195e-08 & 7.111e-13 & 1.066e-08 & 1.397e-08 & 3.573e-13 & 7.884e-09\\

50 & 1.140e-08 & 6.181e-13 & 1.079e-08 & 1.352e-08 & 3.851e-13 & 7.790e-09\\

55 & 1.291e-08 & 6.634e-13 & 1.048e-08 & 1.444e-08 & 3.698e-13 & 7.713e-09\\

60 & 1.141e-08 & 6.480e-13 & 9.967e-09 & 1.753e-08 & 3.729e-13 & 7.726e-09\\

65 & 1.001e-08 & 5.310e-13 & 8.971e-09 & 1.154e-08 & 3.776e-13 & 7.395e-09\\

70 & 1.199e-08 & 5.383e-13 & 1.032e-08 & 1.566e-08 & 3.565e-13 & 8.173e-09\\

75 & 1.177e-08 & 5.721e-13 & 1.009e-08 & 1.377e-08 & 3.485e-13 & 7.845e-09\\

80 & 1.264e-08 & 6.803e-13 & 1.331e-08 & 1.358e-08 & 3.917e-13 & 8.012e-09\\

85 & 1.326e-08 & 6.928e-13 & 1.008e-08 & 1.717e-08 & 3.655e-13 & 7.591e-09\\

90 & 1.315e-08 & 6.055e-13 & 1.012e-08 & 1.746e-08 & 3.629e-13 & 8.078e-09\\

95 & 1.174e-08 & 4.775e-13 & 1.118e-08 & 1.616e-08 & 3.637e-13 & 8.059e-09\\

100 & 1.018e-08 & 8.551e-13 & 1.065e-08 & 1.404e-08 & 3.604e-13 & 7.793e-09\\
\bottomrule
\end{tabular}
\label{tab:res_vnodes_glm}
\end{table}

\section{Datasets}
\label{sec:a:datasets}

\subsection{Diamonds}
\begin{figure}[t]
\centering
\includegraphics[width=\textwidth]{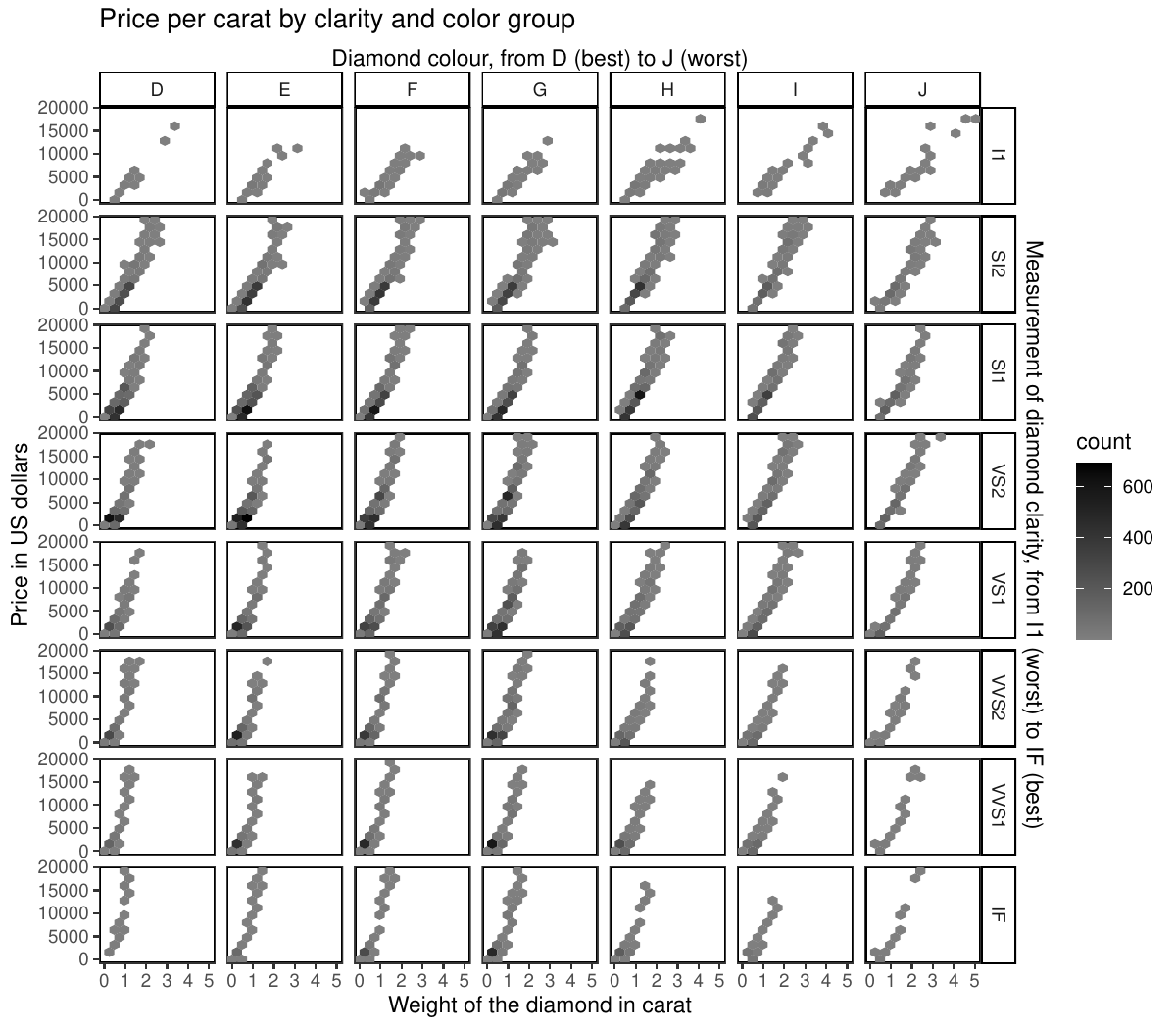}
\caption{Scatterplot of the relation of price and carat in a hexagon bin count format, with the number of bins set to 10, and each combination of the color and clarity variables}
\label{fig:diamonds_facet}
\end{figure}

Figure \ref{fig:diamonds_facet} presents a scatterplot of the relation of the price of a diamond and the weight in carat with a hexagon bin count format --- where the number of bins was set to 10 --- and each combination of the categorical variables color and clarity.
From the figure, it's possible to see an upwards trend in the diamond price when both the colour and clarity are considered of a higher grade.
It is particular useful to look at the first figure from the the last line --- best diamond colour D and best measurement of clarity IF --- where the price increases much faster with small increases in weight.

\end{document}